\documentclass[journal]{IEEEtran}

\usepackage{amsmath}
\usepackage{tabulary}
\usepackage{booktabs}
\usepackage{units}
\usepackage{multirow}
\usepackage{subfig}
\usepackage{graphicx}
\usepackage{gensymb}
\usepackage{url}

\begin{document}
\title{Vectorized Calculation of Short Circuit Currents Considering Distributed Generation - An Open Source Implementation of IEC 60909}
\author{Leon~Thurner,~Martin~Braun
\thanks{L. Thurner is with the Department of Department of Energy Management and Power System Operation, University of Kassel, 34121, Germany, e-mail: leon.thurner@uni-kassel.de
M. Braun is with the Department of Energy Management and Power System Operation, University of Kassel and the Fraunhofer IEE, 34121 Kassel, Germany }}

\markboth{}%
{}
\maketitle

\begin{abstract}
An important task in grid planning is to ensure that faults in the grid are detected and cut off without damage in any grid element. Calculating short circuit currents is therefore a vital grid analysis functionality for grid planning applications. The standard IEC 60909 provides guidelines for short circuit calculations and is routinely applied in grid planning applications. This paper presents a method for the vectorized calculation of short circuit currents according to IEC 60909.  Distributed generation units are considered according to the latest revision of the standard. The method is implemented in the python based open source tool \texttt{pandapower} and validated against commercial software and examples from literature. The implementation presented in this paper is the first comprehensive implementation of the IEC 60909 standard which is available under an open source license. It can be used to evaluate fault currents in grid studies with a high degree of automation and is shown to scale well for large grids. Its practical applicability is shown in a case study with a real MV grid with a high degree of DG penetration. \end{abstract}


\IEEEpeerreviewmaketitle
\section{Introduction}
\IEEEPARstart{F}{aults} in electric grids occur due to external interference, such as a tree falling on an overhead line, or due to failures in grid elements, such as a disruptive discharge in a cable due to a failure in the insulation material. These faults lead to transient short circuit (SC) currents that can be higher than normal operational currents by several magnitudes. These currents impose a high thermal as well as mechanical strain on lines, transformers and other power system components. Electrical grids have to be designed so that SC currents are limited and faults can not lead to melting, combustion or even explosion in any grid components. It also has to be ensured that all faults can be detected and cut off by the protection systems. Calculating SC currents is therefore a crucial analysis functionality in grid planning and protection system design.

\subsection{IEC 60909}
\begin{figure}[b]
\centering
\includegraphics[width=.45\textwidth]{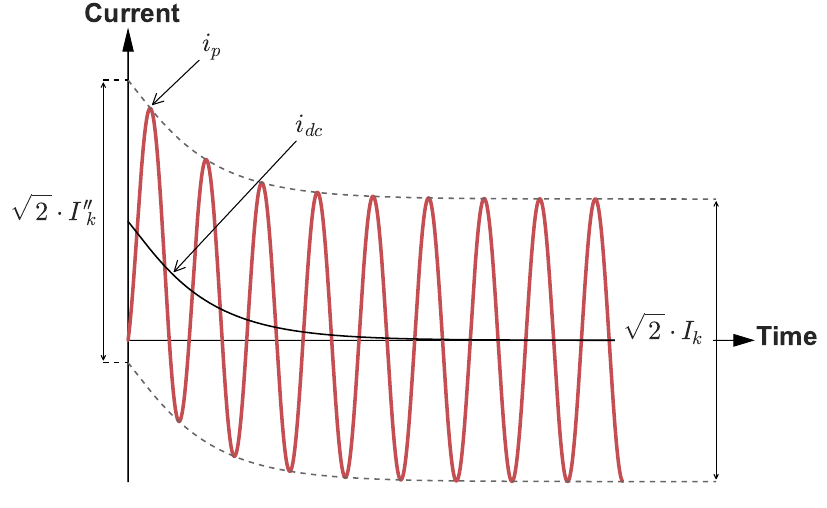}%
\caption{Characteristic curve of a SC current far from a generator}
\label{fig:sc_curve}
\end{figure}

Faults lead to transient currents that can be simulated with dynamic methods. In grid planning however, many possible fault scenarios have to be considered in advance, which makes detailed dynamic simulations of each fault infeasible. To allow approximations of SC currents with static methods, the Technical Committee 73 of the IEC has put forward the standard IEC 60909~\cite{iec60909}. The standard characterizes the SC curve with an initial SC current $I^{''}_k$, peak short circuit current $i_p$ or long term SC current $I_k$ as shown in Figure~\ref{fig:sc_curve}. Detailed instructions for how these currents can be calculated from the static grid model are provided in the standard. To account for transient effects as best as possible with static calculation methods, the standard defines multiple correction factors and other calculation rules. The most recent revision of the standard also defines a method on how to integrate the contribution of distributed generation (DG) to the SC current, which have previously been neglected~\cite{balzer}. The IEC 60909 standard is widely applied in grid planning applications~\cite{Nippert2005, Sweeting2012, Karaliolios2008, Boljevic2009, Tanaka}.

\subsection{Available Tools}
SC calculations according to IEC 60909 is a standard functionality in commercial power system analysis tools like Neplan\footnote{http://www.neplan.ch/}, PSS Sincal\footnote{http://w3.siemens.com/smartgrid/global/en/products-systems-solutions} or DIgSILENT PowerFactory\footnote{http://www.digsilent.com}. These tools can be used by grid planners and are mostly designed as graphical user interface (GUI) applications where single calculations are triggered manually. For scientific purposes however, it is often necessary to automate the analysis of grids. It is also desirably to use open source software that does not require license fees and allows unconstrained parallel computing. While many open source tools exist that allow automation of power flow and optimal power flow evaluations, SC calculations are only supported by few tools. Generic SC calculations without considering the specifications and correction factors defined in the standard are supported by GridCal\footnote{https://github.com/SanPen/GridCal}, InterPSS\footnote{www.interpss.com} and OpenDSS\footnote{http://smartgrid.epri.com/SimulationTool.aspx}. SC calculations according to IEC 60909 are considered in the open source tool Elplek\footnote{http://pp.kpnet.fi/ijl/}, which also implements functionality for protection system studies. Elplek is however a GUI application with a focus on manual single case evaluations and is neither suited for automated analysis nor does it scale well for large grids. There is currently no open source tool known to the authors that allows vectorized SC calculations according to IEC 60909 including the latest revision for the consideration of DG. This gap is closed by the implementation described in this paper. It is available in the open source tool \texttt{pandapower}~\cite{pandapower}.
This paper is structured as follows: Section \ref{sec:pandapower} describes the open source tool \texttt{pandapower} and how it can be used to implement SC calculations. Section \ref{sec:grid_model} describes how the different grid elements are modelled for the fault case, and Section \ref{sec:sc_currents} describes how the SC currents are calculated based on the grid model. Section \ref{sec:case_study} shows the application of the SC calculation in a case study from literature as well as a real MV grid. Section \ref{sec:conclusion} provides a summary and a conclusion of this paper.

\section{pandapower} \label{sec:pandapower}
This paper describes the implementation of a vectorized SC calculation considering the specifications of IEC 60909 in the grid analysis tool \texttt{pandapower}.

\subsection{About \texttt{pandapower}}
\texttt{pandapower} is a BSD licensed open source module implemented in Python with a data structure based on the data analysis library pandas~\cite{pandapower}. It is an easy to use grid calculation and analysis framework that includes equivalent circuit models for elements such as lines, two and three-winding-transformers, loads, external grids, synchronous generators and more~\cite{pandapower}. Electric elements are specified with nameplate data, such as short circuit voltages and rated apparent power for transformers or length of lines and relative resistances in Ohm. \texttt{pandapower} also includes a switch model, a feature which is especially important in radial systems. All element models are thoroughly tested and validated against commercial software. On the basis of this grid model, \texttt{pandapower} offers power flow, optimal power flow, state estimation and topological graph search functions. The implementation of SC calculations according to IEC 60909 presented in this paper further extends the comprehensive grid analysis functionality of the module.

\subsection{Grid Representation} \label{sec:grid_model}
\begin{figure}
\centering
\subfloat[Bus-branch grid model]{\includegraphics[width=.4\textwidth]{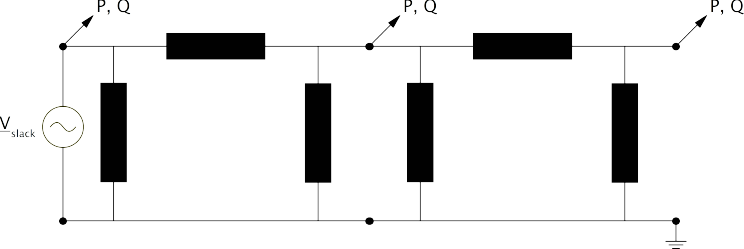} \label{fig:bbm}} 

\subfloat[Element based grid model]{\includegraphics[width=.4\textwidth]{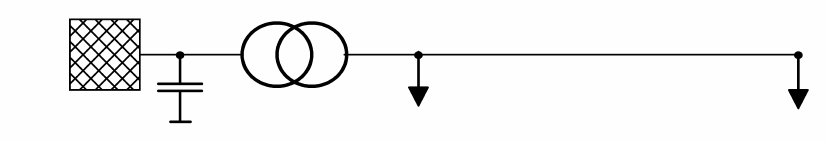} \label{fig:element_based}}%
\caption{Different grid representations}
\label{fig:grid_model}
\end{figure}

To carry out a SC calculation, the grid parameters of all grid elements have to be defined by the user. There are different possibilities of how this input parameters can be provided. A common approach is the bus-branch model (BBM), which defines the grid as a generic collection of buses which are connected by branches~\cite{matpower} (see Figure~\ref{fig:bbm}). The branches are modelled with a predefined equivalent circuit and can be used to model lines or transformers. However, all branch impedances and summed power injections have to be calculated from the nameplate data of the grid elements by the user.
Instead of a generic branch model, \texttt{pandapower} uses an element based model (EBM) with separate models for lines, two-winding and three-winding transformers. All element models are internally processed with appropriate equivalent circuits to model the behaviour of the respective element. Internally, all elements are combined into a BBM that can be used for grid analysis functions. This also has the advantage that different equivalent circuits can be specified depending on the grid analysis function that is being used. For example, \texttt{pandapower} provides different transformer models for power flow and SC calculation. The transformer model is defined with the same nameplate parameters by the user, but is internally modelled with a SC equivalent circuit for SC calculations and a T-model for power flow calculations. Since the definition of element parameters is decoupled from the circuit impedances, it is also possible to include correction factors defined for elements in the SC calculation. This would have to be done manually by the user if a BBM was used.

\section{Short Circuit Grid Model} \label{sec:grid_model}
The grid is defined by the user with an EBM as described in section~\ref{sec:grid_model}. When a SC calculation is carried out, the EBM is converted into a BBM. The equivalent circuit models including specified correction factors are considered as defined in the standard automatically. The BBM is mathematically represented in a nodal point admittance matrix $\underline{Y}$ that includes the impedances of all grid elements~\cite{Grainger1994}. This section introduces the most important models for distribution systems analysis, all available models can be found in the \texttt{pandapower} documentation\footnote{https://pandapower.readthedocs.io}.

\subsection{Bus Elements}
There are two different kinds of bus elements which are considered in the short circuit calculation:
\begin{itemize}
\item Electric motors or generators, which are modelled as a voltage source with internal impedance
\item Full converter elements, which are modelled with a current source according to the 2016 revision of the standard 
\end{itemize}

Constant power loads and shunt elements are neglected according to the standard.

\subsubsection{Voltage Source Elements}
The internal impedances of voltage source elements can be calculated from their nameplate data. The relevant formulas are provided in the standard.
The internal impedance of an external grid is calculated as~\cite{iec60909}:
\begin{equation}
Z_{Q} = \frac{c \cdot U_{nQ}}{\sqrt{3} I''_{kQ}} = \frac{c \cdot U_{nQ}^2}{S''_{kQ}}
\end{equation}
where $S''_{kQ}$ is the SC power of the grid and $c$ is the voltage correction factor. Since external grid connections are equivalent elements representing aggregated upstream grid groups with multiple generators, the SC power has to be retrieved from measurements or grid reduction. To account for worst case situations, two values $S''_{kQ, min}$/$S''_{kQ, max}$ are given as well as minimum and maximum values for the $R/X$ ratio of the grid $RX_Q$ to calculate the complex grid impedance. The voltage correction factor $c$ accounts for operational deviations from the nominal voltage in the grid. The standard defines different correction factors which represent worst-case values for minimum and maximum SC calculations as shown in TABLE~\ref{tab:c}.

\begin{table}[b]
\centering
\caption{Voltage correction factor c according to~\cite{iec60909} \label{tab:c}} 
\begin{tabular}{l c c c c}
Voltage Level  &  Tolerance  & $c_{min}$ & $c_{max}$   \\ \toprule
$\multirow{2}{*}{\unit[0.1]{kV} - \unit[1]{kV}}$ & \unit[6]{\%} & 0.95  & 1.05 \\
& \unit[10]{\%} & 0.95 & 1.10 \\ 
$> \unit[1]{kV}$ &  & 1.00 & 1.10  \\ \midrule
\end{tabular}
\end{table}

\subsubsection{Current Source Elements}
Full converter elements, e.g. PV or wind power plants, are modelled as current sources. The current injection by DG is assumed to be inductive, so that the current is calculated as~\cite{balzer}:
\begin{equation}
\underline{I}_{kU} = - j (k \cdot I_{rU})
\end{equation}
where the rated current $I_{rU}$ and the ratio of short circuit to rated current $k$ are given by the manufacturer.

\subsection{Branch Elements}
\begin{figure}
\centering
\subfloat[Line and Two-Winding Transformer]{\includegraphics[width=.19\textwidth]{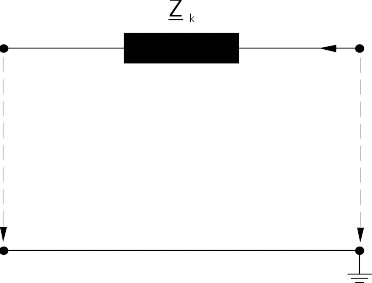} \label{fig:two_winding}} \qquad \qquad
\subfloat[Three-Winding Transformer]{\includegraphics[width=.11\textwidth]{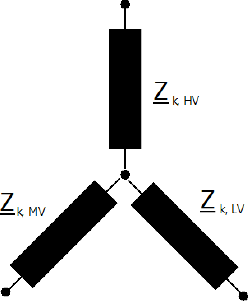} \label{fig:three_winding}}%
\caption{Equivalent circuits for branch elements}
\label{fig:branches}
\end{figure}

Lines and two-winding transformers are represented by a single SC impedance as shown in Figure~\ref{fig:two_winding}, and three winding transformers are represented by a star equivalent circuit as shown in Figure~\ref{fig:three_winding}. Shunt admittances of all branch elements are neglected.

\subsubsection{Line}
The short circuit impedance of a line is equal to the normal operation line impedance for maximum SC current calculations. For minimum SC currents, the line resistance at standard operation temperature $R_{L20}$ is corrected in relation to the end temperature of the conductor $\theta_e$ after a fault~\cite{iec60909}:
\begin{equation}
R_L = [1 + \unit[0.04]{K^{-1}} (\theta_e - \unit[20]{\degree C})] \cdot R_{L20}
\end{equation}

\subsubsection{Two-Winding Transformer}
The relative transformer impedance $z_k$ can be calculated from the short circuit voltage $u_k$ as:
\begin{equation}
z_k = \frac{u_k}{100} \cdot K_T
\end{equation}
with the correction factor $K_T$~\cite{iec60909}:
\begin{equation}
K_T = 0.95 \cdot \frac{c_{max}}{1 + 0.6 \cdot x_T} \label{eq:kt}
\end{equation}
where $c_{max}$ is the maximum voltage correction factor on the low voltage side of the transformer and $x_T$ is the transformer impedance relative to the rated values of the transformer.

\subsubsection{Three-Winding Transformer}
Three winding transformers are modelled by three equivalent two-winding transformers~\cite{BrownBook}. The three equivalent two-winding transformers are represented by their SC impedance in a star connection as shown in Figure~\ref{fig:three_winding}. The transformer correction factor given in equation~\ref{eq:kt} is also applied to the equivalent two-winding transformers.

\section{Short-Circuit Current Calculation} \label{sec:sc_currents}
\begin{figure}
\centering
\subfloat[3-bus example grid]{\includegraphics[width=.4\textwidth]{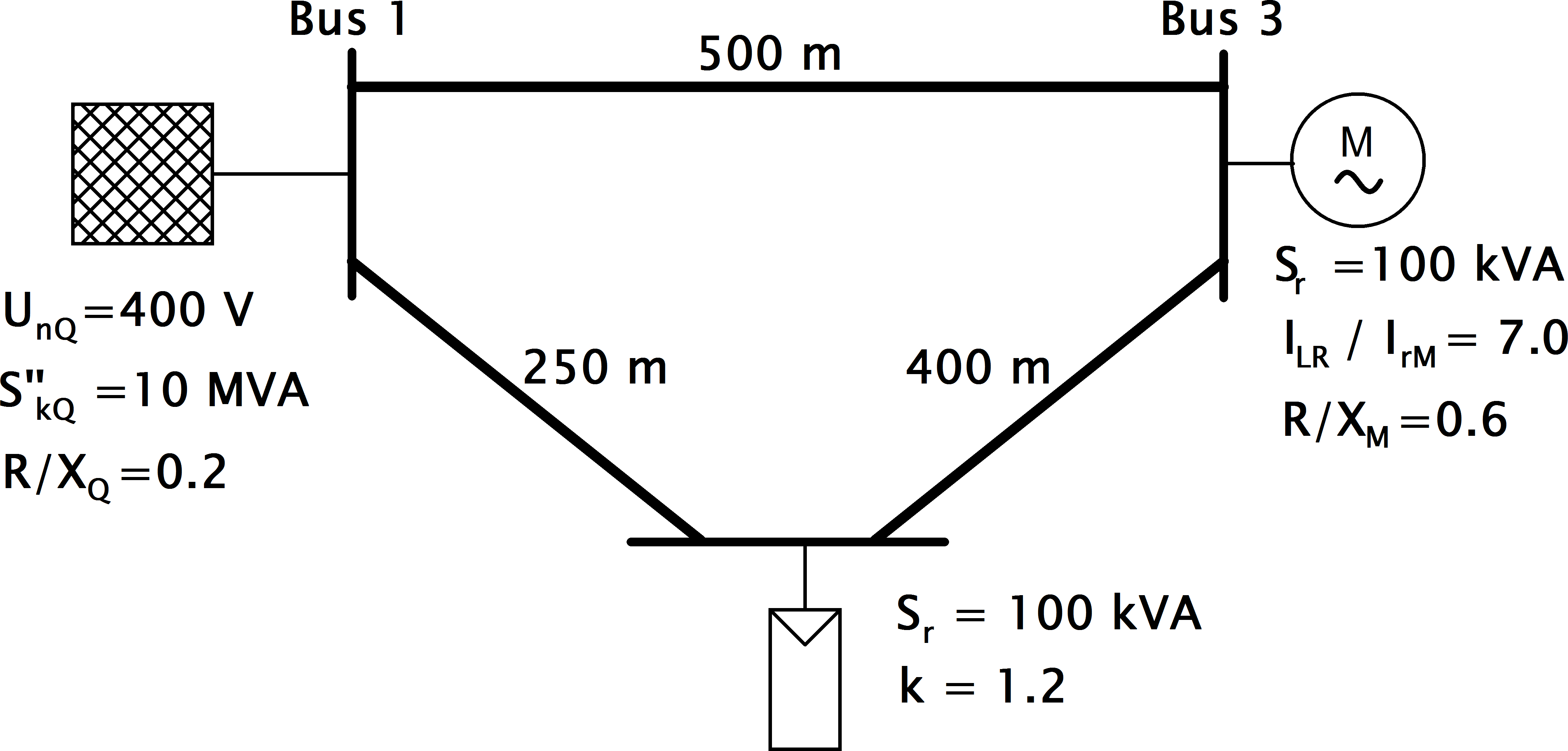} \label{fig:3bus}}

\subfloat[Equivalent circuit for fault at bus 3]{\includegraphics[width=.42\textwidth]{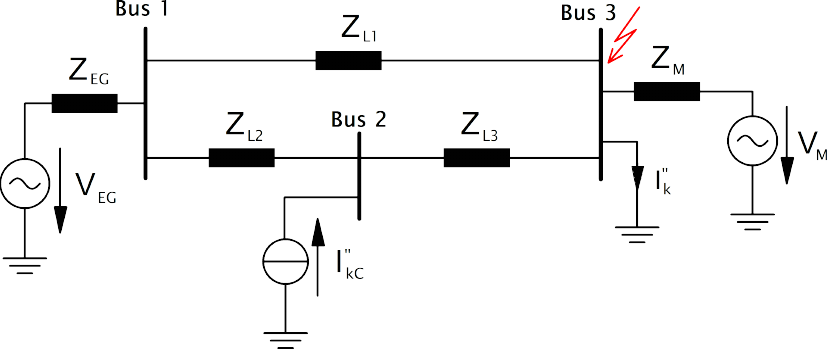} \label{fig:3bus_equivalent}}%

\subfloat[Equivalent voltage source at fault location]{\includegraphics[width=.42\textwidth]{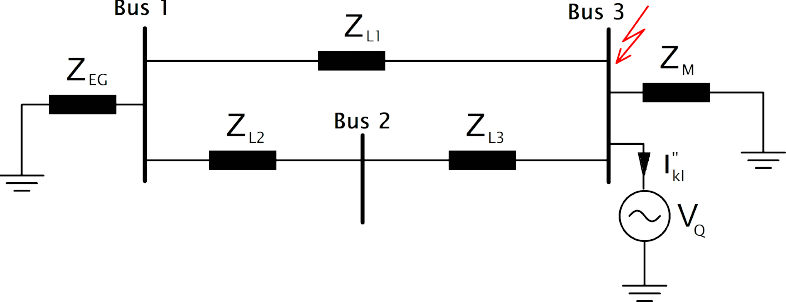} \label{fig:3bus_voltage}}%

\subfloat[Equivalent circuit for consideration of current sources]{\includegraphics[width=.42\textwidth]{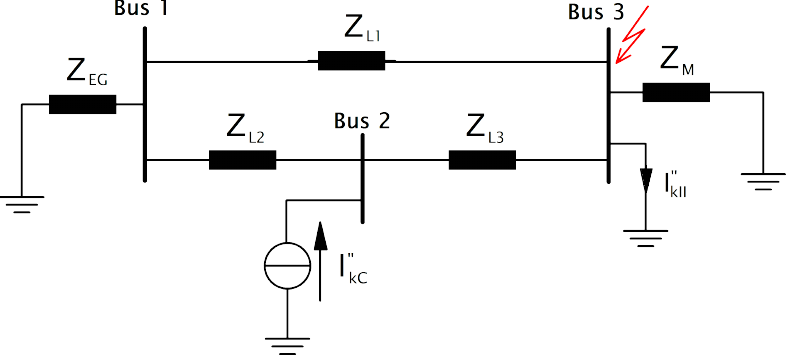} \label{fig:3bus_current}}%

\caption{Example grid with equivalent circuits}
\label{fig:bus_sources}
\end{figure}

The SC current is calculated in three steps:
\begin{itemize}
\item calculate the SC contribution $I''_{kI}$ of all voltage source elements
\item calculate the SC contribution $I''_{kII}$ of all current source elements
\item calculate the total initial SC current $I''_k = I''_{kI} + I''_{kII}$ 
\end{itemize}
This process is shown step by step for the three-bus example grid. The example grid is shown in Figure ~\ref{fig:3bus} in normal operation. The equivalent circuit for a short circuit at Bus 3 is shown in Figure~\ref{fig:3bus_equivalent}. The circuit is then separated into two equivalent circuits for the calculation of $I''_{kI}$ (Figure~\ref{fig:3bus_voltage}) and $I''_{kII}$ (Figure~\ref{fig:3bus_current}). The detailed steps of how these equivalent circuits are built and evaluated to calculate I are discussed in this section.

\subsection{Equivalent Voltage Source}
In the first step, the voltage source contribution to the SC current is calculated. The current source elements are therefore neglected. Furthermore, all voltage sources are replaced with one equivalent voltage source at the fault location according to the theorem of Thevenin. The magnitude of the equivalent voltage source is given as~\cite{iec60909}:
\begin{equation}
U_Q = \frac{c \cdot U_N}{\sqrt{3}}
\end{equation}
where $U_N$ is the nominal voltage and c is either $c_{min}$ for minimum SC calculations or $c_{max}$ for maximum SC calculations. Since all load currents can be neglected for SC calculations, the currents at all buses are zero except for the fault bus, where the current is equal to the SC current. The grid equations for a fault at bus $j$ can then be derived from the ohmic law as:

\begin{equation}
    \begin{bmatrix}
    \underline{U}_{1}  \\
    \vdots  \\[0.4em]
    \underline{U}_{Qj}  \\[0.4em]
    \vdots  \\
    \underline{U}_{n}
    \end{bmatrix}  
    = 
    \begin{bmatrix}
    \underline{Z}_{11} & \dots & \dots & \dots & \underline{Z}_{n1} \\
    \vdots & \ddots &  & & \vdots \\
    \vdots & & \underline{Z}_{jj} & & \vdots \\
    \vdots & & & \ddots & \vdots \\
    \underline{Z}_{1n} & \dots & \dots & \dots & \underline{Z}_{nn}
    \end{bmatrix}
    \begin{bmatrix}
    0 \\
    \vdots  \\[0.25em]
    \underline{I}''_{kIj} \\[0.25em]
    \vdots  \\
    0 
    \end{bmatrix}
\end{equation}
\\[0.5em]
where the impedance matrix $\underline{Z}$ is the inverse of the nodal point admittance matrix $\underline{Y}$. The SC current at fault bus $j$ can now be extracted from row $j$ of the matrix equation as:
\begin{equation}
\underline{I}''_{kIj} = \frac{\underline{U}_{Qj}}{\underline{Z}_{ij}}
\end{equation}
To calculate the vector of the SC currents at all buses, the matrix equation can be expanded as follows:
\begin{equation*}
    \begin{bmatrix}
    \underline{U}_{Q1} & \dots & \underline{U}_{n1} \\[.8em]
    \vdots & \ddots & \vdots \\[.8em]
    \underline{U}_{1n} & \dots & \underline{U}_{Qn}
    \end{bmatrix}  
    = 
    \begin{bmatrix}
    \underline{Z}_{11} & \dots & \underline{Z}_{n1} \\[0.8em]
    \vdots & \ddots & \vdots \\[0.8em]
    \underline{Z}_{1n} & \dots & \underline{Z}_{nn}
    \end{bmatrix}
    \begin{bmatrix}
    \underline{I}''_{kI1} & \dots & 0 \\[0.8em]
    \vdots & \ddots & \vdots \\[0.8em]
    0 & \dots & \underline{I}''_{kIn}
    \end{bmatrix}
\end{equation*}
Since the current matrix is diagonal, the vector of SC current magnitudes at all buses can be calculated as:
\begin{equation}
    \begin{bmatrix}
    I''_{kI1} \\[0.25em]
    \vdots  \\[0.25em]
    I''_{kIn} \\
    \end{bmatrix}
    = 
    \begin{bmatrix}
    \frac{U_{Q1}}{Z_{11}}  \\
    \vdots  \\
    \frac{U_{Qn}}{Z_{nn}} 
    \end{bmatrix}
\end{equation}

\subsection{Contribution of Current Source Elements}
To calculate the current source component of the SC current, all voltage sources are short circuited and only current sources are considered. The bus currents are then given as:

\begin{equation} \label{eq:ikc}
    \begin{bmatrix}
    \underline{I}_1 \\[0.2em]
    \vdots  \\[0.2em]
    \underline{I}_m \\[0.2em]
    \vdots  \\
    \underline{I}_n
    \end{bmatrix}
    =
    \begin{bmatrix}
    0 \\[0.2em]
    \vdots  \\[0.2em]
    \underline{I}''_{kIIj} \\[0.2em]
    \vdots  \\
    0
    \end{bmatrix}
    -
    \begin{bmatrix}
    \underline{I}''_{kC1} \\[0.2em]
    \vdots  \\[0.2em]
    \underline{I}''_{kCj} \\[0.2em]
    \vdots  \\
    \underline{I}''_{kCn}
    \end{bmatrix}
    =
    \begin{bmatrix}
    -\underline{I}''_{kC1} \\[0.2em]
    \vdots  \\[0.2em]
    \underline{I}''_{kIIj} - \underline{I}''_{kCj} \\[0.2em]
    \vdots  \\
    -\underline{I}''_{kCn}
    \end{bmatrix}
\end{equation}
where $I''_{kC}$ are the SC currents that are fed in by the converter element at each bus and $I''_{IIj}$ is the contribution of converter elements at the fault bus j. With the voltage at the fault bus known to be zero, the network equations are given as:
\begin{equation}
    \begin{bmatrix}
    \underline{U}_{1}  \\
    \vdots  \\[0.4em]
    0  \\[0.4em]
    \vdots  \\
    \underline{U}_{n}
    \end{bmatrix}  
    = 
    \begin{bmatrix}
    \underline{Z}_{11} & \dots & \dots & \dots & \underline{Z}_{n1} \\
    \vdots & \ddots &  & & \vdots \\
    \vdots & & {Z}_{jj} & & \vdots \\
    \vdots & & & \ddots & \vdots \\
    \underline{Z}_{1n} & \dots & \dots & \dots & \underline{Z}_{nn}
    \end{bmatrix}
    \begin{bmatrix}
    -I''_{kC1} \\[0.2em]
    \vdots  \\[0.2em]
    \underline{I}''_{kIIj} - \underline{I}''_{kCj} \\[0.2em]
    \vdots  \\
    -I''_{kCn}
    \end{bmatrix}
\end{equation}
 
From which row j of the equation yields:
\begin{equation}
    0 = \underline{Z}_{jj} \cdot \underline{I}''_{kIIj} - \sum_{m=1}^{n}{\underline{Z}_{jm} \cdot \underline{I}_{kCj}}
\end{equation}

which can be converted into:

\begin{equation}
\underline{I}''_{kIIj} = \frac{1}{\underline{Z}_{jj}} \cdot \sum_{m=1}^{n}{\underline{Z}_{jm} \cdot \underline{I}_{kC, m}}
\end{equation}

To calculate all SC currents for faults at each bus simultaneously, this can be generalized into the following matrix equation:
\begin{equation}
    \begin{bmatrix}
    \underline{I}''_{kII1} \\[0.5em]
    \vdots  \\[0.5em]
    \vdots  \\[0.5em]
    \underline{I}''_{kIIn}
    \end{bmatrix} = 
    \begin{bmatrix}
    \underline{Z}_{11} & \dots & \dots & \underline{Z}_{n1} \\[0.3em]
    \vdots & \ddots & & \vdots \\[0.3em]
    \vdots & & \ddots & \vdots \\[0.3em]
    \underline{Z}_{1n} & \dots & \dots & \underline{Z}_{nn}
    \end{bmatrix}
    \begin{bmatrix}
    \frac{I''_{kC1}}{\underline{Z}_{11}} \\[0.25em]
    \vdots  \\
    \vdots  \\[0.25em]
    \frac{I''_{kCn}}{\underline{Z}_{nn}}
    \end{bmatrix}
\end{equation}

\begin{figure}[t]
\centering
\subfloat[HV example network with wind parks]{\includegraphics[width=.45\textwidth]{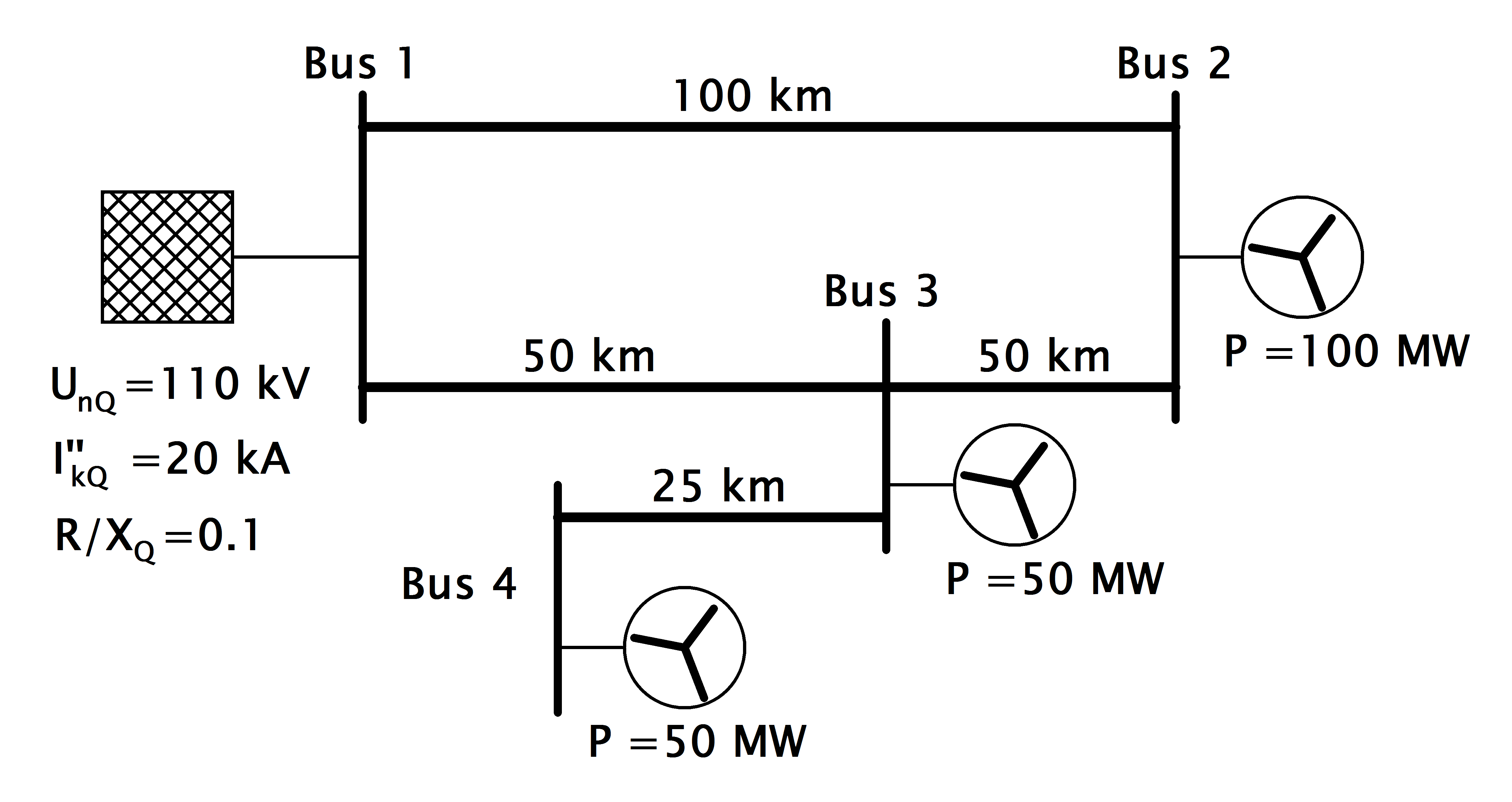} \label{fig:balzer}}

\subfloat[\texttt{pandapower} results for validation of example HV network]{\includegraphics[width=.4\textwidth]{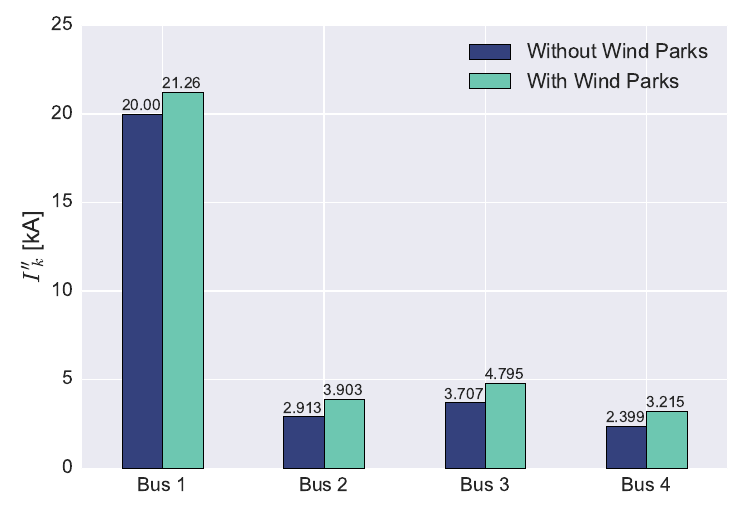} \label{fig:balzer_results}}%
\caption{Validation of SC implementation with three wind parks example from~\cite{balzer} }
\end{figure}

\subsection{Example Network}
For the example network shown in Figure~\ref{fig:3bus}, a calculation of the maximum initial SC current with a voltage tolerance of 10\% yields the following result:

\begin{equation*}
I''_k =  \begin{bmatrix} 145.073 \\ 3.844 \\ 4.524\end{bmatrix} kA + \begin{bmatrix} 0.181 \\ 0.208 \\ 0.117\end{bmatrix} kA = \begin{bmatrix} 145.254 \\  4.052 \\ 4.641\end{bmatrix}  kA
\end{equation*}
The detailed derivation of these currents including the definition of the nodal point admittance matrix is given in Appendix A.

\section{Examples and Case Studies} \label{sec:case_study}
This section includes some exemplary case studies to demonstrate how the \texttt{pandapower} SC module can be used.

\subsection{Impact of Wind Parks in a High Voltage Network}
The example from~\cite{balzer} is used to validate the implementation of the SC calculation with full converter elements. According to~\cite{balzer}, a fault at Bus 2 in the network depicted in Figure~\ref{fig:balzer} with $k=1.2$ leads to a SC current of:
\begin{equation*}
I''_{k2} = \unit[2.913]{kA} + \unit[0.990]{kA} = \unit[3.903]{kA}
\end{equation*}
To validate the implementation presented in this paper, a SC calculation with and without wind parks is run with \texttt{pandapower}. The results can be seen in Figure~\ref{fig:balzer_results} and are consistent with the results given in~\cite{balzer}. Because of the vectorized implementation, \texttt{pandapower} also returns the SC currents at all other buses.

\begin{figure}
\centering
\includegraphics[width=.48\textwidth]{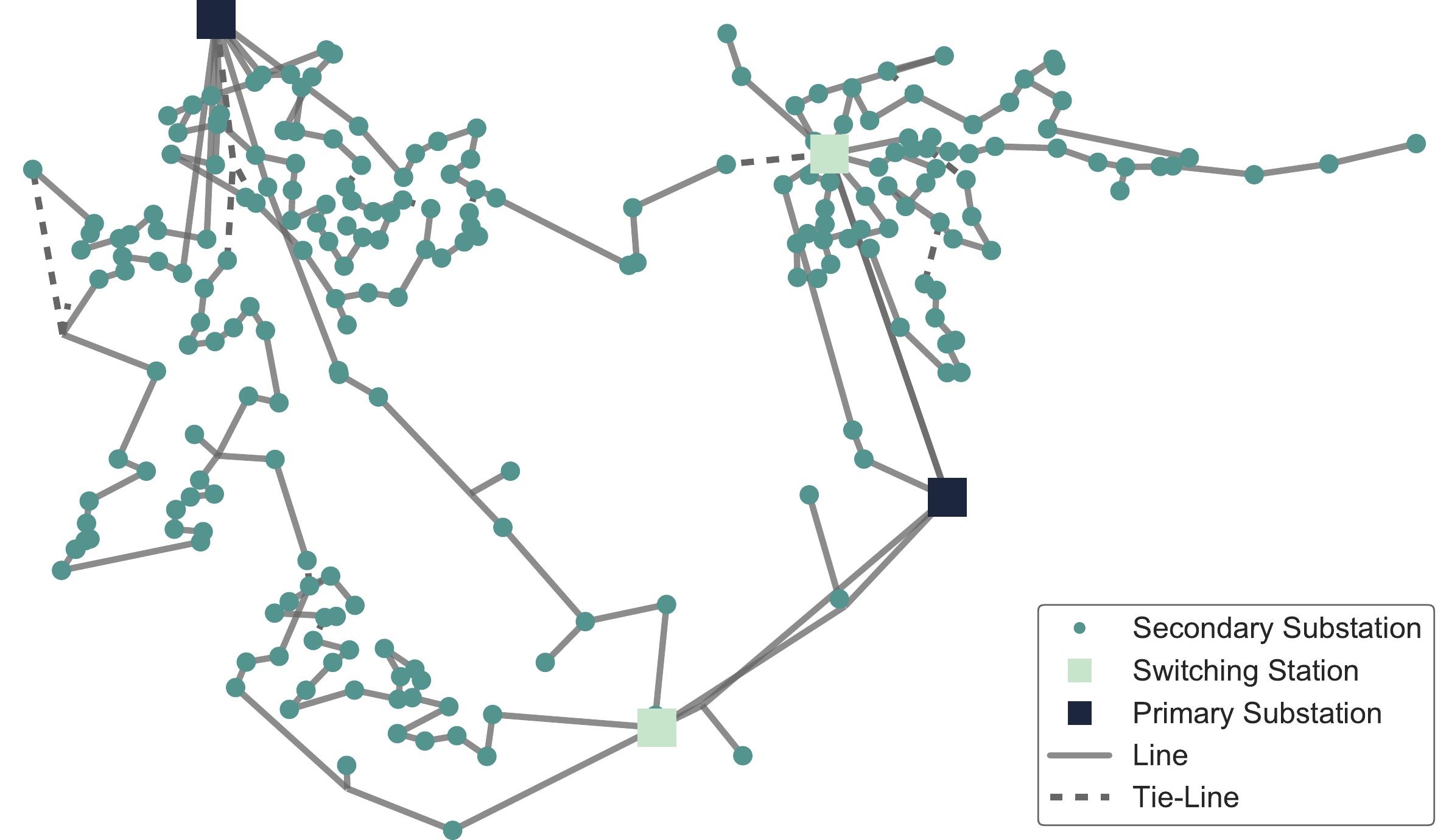}
\caption{Real MV example grid \label{fig:mv_example}}
\end{figure}

\subsection{MV Network Case Study}
\begin{figure}
\centering
\subfloat[Maximum Initial Short Circuit Currents at all buses]{\includegraphics[width=.5\textwidth]{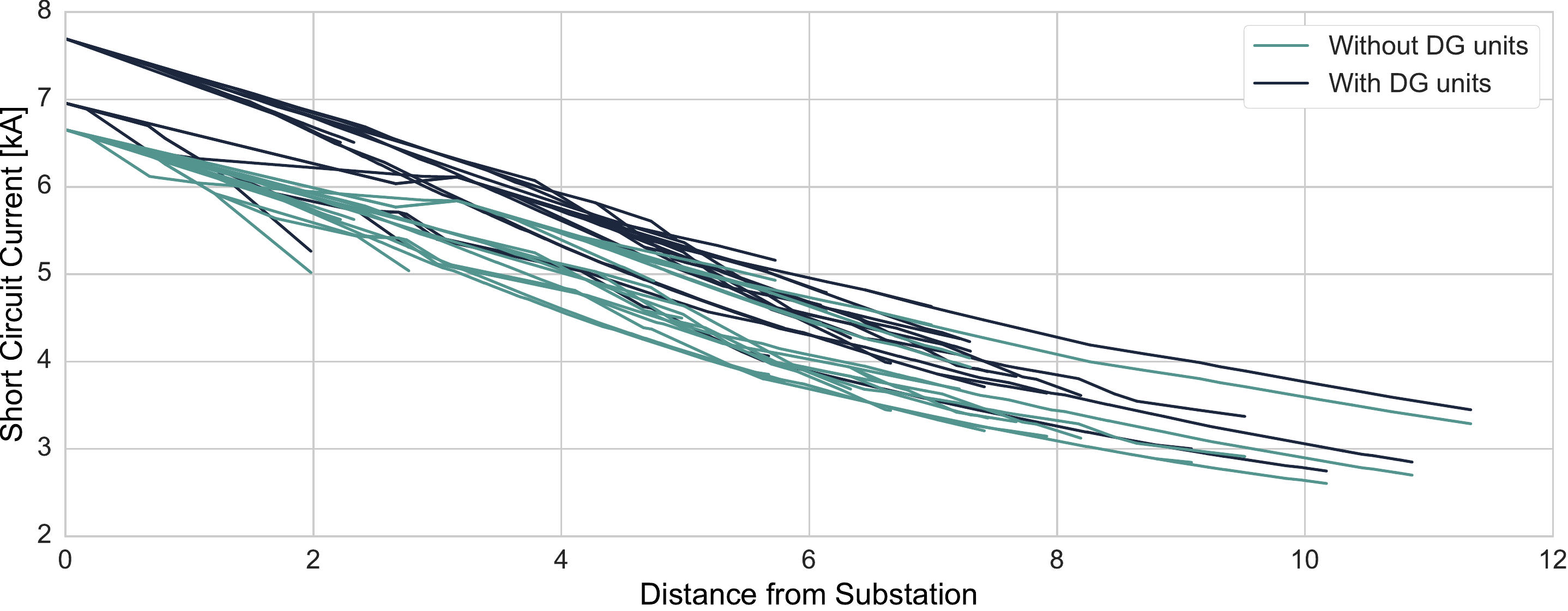} \label{fig:example_max}}%

\subfloat[Minimum Initial Short Circuit Currents at all buses]{\includegraphics[width=.5\textwidth]{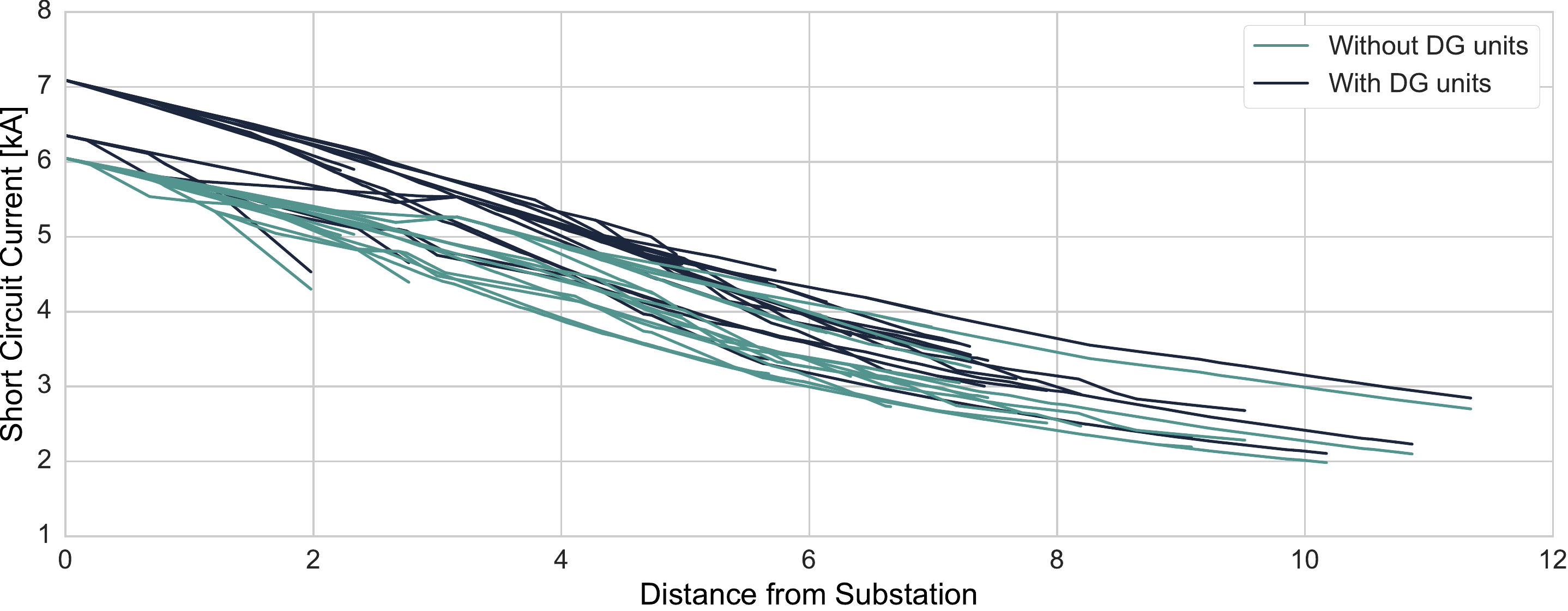} \label{fig:example_min}}
\caption{Real MV example grid \label{fig:mv_example_results}}
\end{figure}

The short circuit calculation is applied in a case study to the real Medium Voltage (MV) distribution grid shown in Figure~\ref{fig:mv_example}. The distribution grid is operated at a nominal voltage of $\unit[20]{kV}$ and is connected to the $\unit[110]{kV}$ high voltage (HV) network through two HV/MV substations. It services 282 MV/LV stations with a combined peak load of about \unit[11]{MW}. In addition, there are DG units with a rated power of over \unit[13]{MW} installed in this grid area. With this high penetration of DG, this area is well suited to demonstrate the impact of DG on the short circuit current in a real use case. Figure~\ref{fig:mv_example_results} shows the maximum (Figure~\ref{fig:example_max}) and minimum (Figure~\ref{fig:example_min}) short circuit currents for faults at all buses. Since the network is operated radially, the impedance between fault and external grid connection increases the larger the distance between fault location and HV/MV substation. That is why the short circuit current decreases with growing distance to the HV/MV substation. The overall maximum short circuit current can therefore be found directly at the HV/MV substation, whereas the minimal short circuit current is found at the end of the feeder. The short circuit contribution of the DG units is considered with $k = 1.0$. When considering DG in the short circuit calculation according to the 2016 revision of the standard, the maximum short circuit current at the primary substation rises from $\unit[6.65]{kA}$ to $\unit[7.69]{kA}$ (see Figure~\ref{fig:example_max}). This represents an increase by \unit[16]{\%}, which shows that considering the short circuit contribution of DG can be relevant when the short-circuit capabilities of bus bars in the primary substations are assessed. It can also be seen that the short circuit current is the same at the beginning of all feeders when DG are not considered, since the short circuit current is fed only by the external grid connection. When DG are considered, the individual feeders show a different behaviour depending on the DG capacity on each feeder, which makes a detailed analysis of the feeders necessary. For the minimum short circuit calculation, the end temperature of the lines after the fault is assumed to be \unit[200]{\degree C} for overhead lines and \unit[90]{\degree C} for underground cables. The minimum short circuit current at the end of the feeder then rises by \unit[6]{\%} from $\unit[1.99]{kA}$ to $\unit[2.11]{kA}$ when considering DG (see Figure~\ref{fig:example_min}). While maximum SC currents are relevant for the SC current capability of the grid components, minimum SC currents are relevant when analysing if all faults can be reliably detected by protection systems.

\section{Conclusion} \label{sec:conclusion}
This paper presents a method for the calculation of the initial short circuit current according to IEC 60909. Full converter elements are considered as current sources according the 2016 revision of the standard. The presented method is implemented in the network analysis tool \texttt{pandapower}, which is available as an open source software. The implementation allows to take lines, two-winding transformers, three-winding transformers, synchronous generators, asynchronous machines and full converter elements into account. While calculation of other short circuit currents like the peak current $i_p$ or the thermal equivalent current $I_{th}$ are not addressed in this paper, they are also available in the \texttt{pandapower} implementation. Results for all elements and short circuit currents are tested and validated against commercial software as well as against results from other publications. The vectorized implementation allows efficient simultaneous calculation of SC currents at all buses even for large networks. This is especially useful in network planning applications, where multiple possible fault scenarios have to be taken into account. Since \texttt{pandapower} is focused on automated evaluations, the presented implementation is well suited to consider SC currents in automated network analysis and optimization studies. The practical applicability of the implementation was demonstrated with a case study of a real MV grid with a high penetration of DG. The case study showed, that DG can have a significant impact on the short circuit currents in distribution systems.

The implementation presented in this paper is the first comprehensive implementation of the IEC 60909 standard in any open source power systems analysis tool. It is therefore a valuable contribution towards closing the gap between commercial and open source power system analysis tools. \texttt{pandapower} is under continuous development on github~\cite{pandapower_github}, and it is planned to add further functionality to the short circuit calculation, such as short circuit currents close to generators or single-phase faults.

\section*{Acknowledgement}
This work was supported by the German Federal Ministry for Economic Affairs and Energy and the Projekttr\"ager J\"ulich GmbH (PTJ) within the framework of the projects \textit{Smart Grid Models} (FKZ: 0325616). The authors furthermore thank Netze BW GmbH for providing grid data as well as helpful feedback and discussions on the topic.

\bibliographystyle{IEEEtran}
\bibliography{bibtex}
\end{document}